\documentclass[pra,aps,amssymb,showpacs,twocolumn]{revtex4}
\usepackage{graphicx}
\newcommand{\ket}[1]{|#1\rangle}
\newcommand{\bra}[1]{\langle#1|}

\newcommand{\mc}{\mathrm}

\begin{document}

\title{Thermodynamical Detection of Entanglement by Maxwell's Demons}
\author{Koji Maruyama$^1$, Fumiaki Morikoshi$^{1,2}$ and Vlatko Vedral$^1$}
 \affiliation{$^1$QOLS, Blackett Laboratory, Imperial College London,
   London SW7 2BW, United Kingdom\\
$^2$NTT Basic Research Laboratories, NTT Corporation, 3-1
Morinosato-Wakamiya, Atsugi, 243-0198, Japan}
\date{\today}

\begin{abstract}
Quantum correlation, or \textit{entanglement}, is now believed to be an indispensable
physical resource for certain tasks in quantum information processing, for which
classically correlated states cannot be useful. Besides information processing, what kind
of \textit{physical} processes can exploit entanglement? In this paper, we show that
there is indeed a more basic relationship between entanglement and its usefulness in
thermodynamics. We derive an inequality showing that we can extract more work out of a
heat bath via entangled systems than via classically correlated ones. We also analyze the
work balance of the process as a heat engine, in connection with the Second Law of
thermodynamics.
\end{abstract}
\maketitle

\section{Introduction}
Entanglement is a form of correlations between two or more quantum systems whose amount
exceeds anything that can be obtained by the laws of classical physics. Bell was first to
clarify the difference between these correlations by measuring statistical regularities
between local measurements on two separated systems
\cite{bell64}. However, even though it is now well established that quantum systems can
be more correlated than classical ones, the central question is whether we can use these
excess correlations to do something useful. There are many indications from the field of
quantum information that entanglement can result in a computational speed up,
nonetheless, there is no precise and proven link between the two at present. In this
paper, we show that there is indeed a more basic relationship between entanglement and
its utility in thermodynamics: We find a particular work-extracting scenario that reveals
the difference between classical and quantum correlations.

There have been some papers on the work-extraction from correlated quantum states. For
example, Refs. \cite{oppenheim02,zurek03} have analyzed the difference between globally
and locally extractable work. It was shown in Ref. \cite{oppenheim02} that a global
observer can extract more work from a pair of quantum systems than two local observers.
In Ref. \cite{zurek03}, a similar difference is discussed in terms of ``discord".

Our goal here is to clarify the difference between classical and quantum correlations,
using locally extractable work from a heat bath via a given state, without comparing it
with globally extractable work. The expected result will be Bell-type inequalities, which
witness entanglement \cite{terhal00}, written with locally observable thermodynamical
quantities. We will present thermodynamical inequalities that are satisfied by all
classically correlated states, but can be violated by entangled states. By classically
correlated states, we mean separable states as in Ref. \cite{werner89}. Our results
suggest a novel connection between the separability of a quantum state and
thermodynamics. Metaphorically, Maxwell's demons can violate the inequalities in an
attempt to break the Second Law of thermodynamics with the excess work due to the
non-classicality of entanglement, although they can never be successful.

\section{Work-extraction scheme}\label{sec_wes}
Suppose that we have a two dimensional classical system, such as a ``one-molecule gas''
which can only be in either the right or the left side of a chamber. If we have full
information about this molecule, i.e., we know its position with certainty, we can
extract $k_B T\ln2$ of work out of a heat bath of temperature $T$ by letting the gas
expand isothermally. If we have only partial information about the system, the
extractable work becomes $k_B T\ln2[1-H(X)]$, where $H(X)$ is the Shannon entropy and $X$
is a binary random variable representing the position of the molecule
\cite{oppenheim02,szilard29,bennett82,vonneumann}.
For simplicity, we set $k_B T\ln2=1$ hereafter so that we can identify the amount of work
with the amount of information in bits.

The same argument can be applied to quantum cases provided that we know the nature of
projection operators $\{P_0, P_1(=P_0^\perp)\}$ employed to obtain the information. The
corresponding Shannon entropy is $H(p)=-p\log_2p-(1-p)\log_2(1-p)$, where
$p=\mc{Tr}(P_0\rho)$ and $\rho$ is the density matrix for the state. In order to extract
work, we can store the measurement results in classical bits so that the same process as
above can be applied. Or, equivalently, we can copy the information about the quantum
state to an ancilla by using controlled-NOT (CNOT) operation with respect to the basis
defined by projectors $\{P_0, P_1\}$, and use this ancilla to extract work after letting
it dephase. A CNOT is a two-bit unitary operation, which flips one of the two bits, the
target bit, if the other, the control bit, is in 1 and does nothing otherwise.

Let us now consider a bipartite correlated system, retained by Alice and Bob. Suppose a
set of identically prepared copies of the system for which Alice chooses
$A_\theta=\{P_\theta,P_\theta^\perp\}$ and Bob chooses
$B_{\theta'}=\{P_{\theta'},P_{\theta'}^\perp\}$ as bases of their measurements with
$\theta (\theta')$ representing the direction of the basis. Alice performs her
measurement with $A_\theta$ and sends all ancillae containing results to Bob. Then, by
analogy with the single molecule case, Bob can extract $1-H(B_{\theta'}|A_\theta)$ bits
of work per pair on his side after compressing the information of his measurement
outcomes, where $H(X|Y)$ is the Shannon entropy of $X$, conditional on the knowledge of
$Y$.

When their system is in a maximally entangled state, such as
$\ket{\Phi^+}=(\ket{00}+\ket{11})/\sqrt{2}$, $H(A_\theta|B_\theta)$ vanishes for all
$\theta$, unlike any other forms of correlation for which $H(A_\theta|B_\theta)$ can take
any value between 0 and 1. This means that we can extract more work from entangled pairs
than from classically correlated pairs.

Figuratively speaking, demons, Alice and Bob, who share entangled states with each other,
can outdo those who have only classically correlated ones in terms of the amount of
extractable work. By demons we mean here any fictitious entities that manipulate
microscopic objects using information available to them, in analogy with Maxwell's demon
\cite{demon}. Their everlasting wish is to violate the Second Law of thermodynamics by
extracting extra work from a heat bath thanks to entanglement, although this attempt
turns out to be thwarted as we will see below.

\begin{figure}
 \begin{center}
  \includegraphics[scale=0.3]{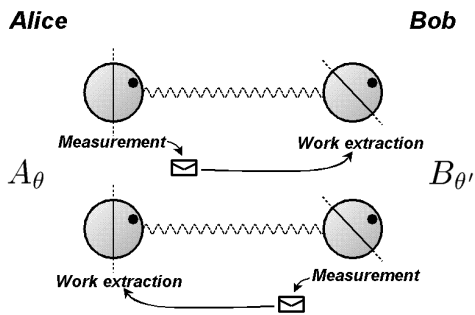}
  \caption{Schematic view of the protocol to extract work from correlated
pairs. Two pairs in the figure represent an ensemble for which
Alice and Bob use $A_\theta$ and $B_{\theta'}$ for their
measurement and work extraction. For a half of this ensemble,
Alice measures her state with $A_\theta$ and Bob extracts work
from his side along the direction of $\theta'$, according to
Alice's measurement results. For the other half, they exchange
their roles.}
 \label{protocol}
 \end{center}
\end{figure}

The overall protocol to extract work from correlated pairs is depicted in Fig.
\ref{protocol}. They first divide their shared ensemble into groups of two pairs to make
the process symmetric with respect to Alice and Bob. For each group, they both choose a
projection operator randomly and independently out of a set, $\{A_1, A_2, \cdots, A_n\}$
for Alice and $\{B_1, B_2, \cdots, B_n\}$ for Bob, just before their measurement. Then,
Alice measures one of the two qubits in a group with the projector she chose and informs
Bob of the outcome as well as her basis choice. Bob performs the same on his qubit of the
other pair in the group. As a result of collective manipulations on the set of those
groups for which they chose $A_i$ and $B_j$, they can extract $2-H(A_i|B_j)-H(B_j|A_i)$
bits of work per two pairs at maximum.

\section{Thermodynamical separability criterion}
\begin{figure}
 \begin{center}
  \includegraphics[scale=0.275]{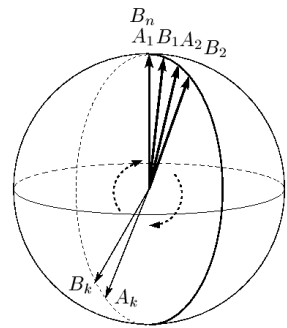}
  \caption{The choice of measurement basis for the protocol.
Bases $A_1, B_1, \cdots, B_n$ are chosen so that they cover a
great circle on the Bloch sphere densely as $n$ tends to infinity.
The final basis $B_n$ is set to be the same as the first one
$A_1$.}
 \label{arrows}
 \end{center}
\end{figure}

Let us find a general description of correlations in terms of work to clarify the
difference between classical and quantum ones. It turns out that it suffices to sum up
all the work that can be obtained by varying the basis continuously over a great circle
on the Bloch sphere, i.e., the circle of maximum possible size on a sphere (see Fig.
\ref{arrows}). This is similar in approach to the chained Bell's inequalities discussed
in Ref. \cite{braunstein90}. The circle should be chosen to maximize the sum.

Let $\xi_\rho(A_i,B_j)$ denote the extractable work from two copies of bipartite system
$\rho$ in the asymptotic limit when Alice chooses $A_i$ (and $A_i^\perp$) as her
measurement and work-extracting basis while Bob chooses $B_j$ (and $B_j^\perp$). For a
two-dimensional bipartite system, $\xi$ is given by
\begin{eqnarray}\label{xidef}
\xi_\rho(A_i,B_j) &:=& 2-H(A_i|B_j)-H(B_j|A_i) \nonumber \\
&=& 2-2H(A_i,B_j)+H(A_i)+H(B_j),
\end{eqnarray}
which is symmetric with respect to $A_i$ and $B_j$.

The quantity we consider is
$\lim_{n\rightarrow\infty}1/(2n-1)[\sum_{k=1}^n\xi(A(\theta_k),
B(\theta_k))+\sum_{k=1}^{n-1}\xi(B(\theta_k),A(\theta_{k+1}))]$ for a state $\rho$ and we
let $\Xi(\rho)$ denote it as
\begin{equation}\label{var_basis}
\Xi(\rho):=\frac{1}{2\pi}\int_0^{2\pi}\xi_\rho(A(\theta),
B(\theta))d\theta,
\end{equation}
where $\theta$ is the angle representing the direction of measurement on the great
circle. The great circle is the one that maximizes the integral, as mentioned above. Note
that $\Xi(\rho)$ represents the extractable work under local operations and classical
communication, which is a standard framework to deal with entanglement in quantum
information theory.

We now present an inequality that shows a connection between the thermodynamically
extractable work and the separability of bipartite quantum systems.

\textit{Proposition.} An inequality
\begin{equation}\label{theineq}
\Xi(\rho)\le\Xi(\ket{00})
\end{equation}
is a necessary condition for a two-dimensional bipartite state $\rho$ to be separable,
that is, $\rho=\sum_i p_i \rho_i^A\otimes\rho_i^B$. The state $\ket{00}$ in the
right-hand side (RHS) can be any pure product state $\ket{\psi\psi'}$. We obtained the
value of $\Xi(\ket{00})$ numerically as 0.8854 bits. We will refer to this inequality
(\ref{theineq}) as ``thermodynamical separability criterion".

\textit{Proof.} Without loss of generality, all $\rho_i^A$ and $\rho_i^B$ can be assumed
as pure states. The key point of the proof is that even if the information from the other
side could always be used to specify the pure state component $\rho_i^B$ (or $\rho_i^A$)
on his/her side, the average extractable work $\Xi(\rho)$ is always not larger than
$\Xi(\ket{00})$. This is because for any pure product state $\ket{\psi\psi'}$,
$\Xi(\ket{\psi\psi'})\le\Xi(\ket{00})$ with equality when both $\ket{\psi}$ and
$\ket{\psi'}$ are on the integral path for $\Xi$. Thus, $\Xi^{pcs}(\rho)=\sum_i p_i
\Xi(\rho_i^A\otimes\rho_i^B) \le \sum_i p_i \Xi(\ket{00})=\Xi(\ket{00})$, where
$\Xi^{pcs}(\rho)$ is the extractable work from $\rho$ when ``pure component specification
(\textit{pcs})" is possible. We show below that $\Xi(\rho)\le\Xi^{pcs}(\rho)$.

The conditional entropy in the definition of $\xi_\rho(A,B)$ can be written as
$H(B|A)=\sum_{j=0}^1 p_j^A H(p(B^0|A^j))$, where $p_j^A=\mc{Tr}[(A^j\otimes
I)\rho]=\sum_i p_i \mc{Tr}(A^j\rho_i^A)$ is the probability for Alice to obtain the
outcome $j$ and $A^j$ is a projection operator for the outcome $j$ along the direction of
$\theta$ \cite{note_ent}. Namely, $A^0(=A^0(\theta))$ corresponds to $A(\theta)$ in Eq. (\ref{var_basis}),
however, we do not write $\theta$ explicitly in this proof for simplicity. As the density
operator for Bob after Alice obtained $j$ is $\rho_{A^j}^B=(1/p_j^A)\sum_i p_i
\mc{Tr}(A^j\rho_i^A)\rho_i^B$, $H(B|A)$ is given by
\begin{equation}\label{condent_real}
H(B|A)=\sum_j p_j^A H\left(\sum_i \frac{p_i}{p_j^A}
\mc{Tr}(A^j\rho_i^A)\mc{Tr}(B^0\rho_i^B)\right).
\end{equation}
On the other hand, if the pcs is possible, the corresponding entropy can be written as
\begin{equation}\label{condent_pcs}
H^{pcs}(B|A)=\sum_i p_i H(\mc{Tr}(B^0\rho_i^B)).
\end{equation}
By letting $p_{ji}^A$ denote $\mc{Tr}(A^j\rho_i^A)$ and noting $p_j^A=\sum_i p_i
p_{ji}^A$, we can have an inequality as
\begin{eqnarray}\label{concavity}
H(B|A) &=& \sum_j p_j^A H\left(\sum_i \frac{p_i
p_{ji}^A}{p_j^A}p_{0i}^B\right) \nonumber \\
&\ge& \sum_i\sum_jp_j^A\frac{p_ip_{ji}^A}{p_j^A}H(p_{0i}^B) \nonumber \\
&=& \sum_i p_i H(p_{0i}^B) =H^{pcs}(B|A),
\end{eqnarray}
due to the concavity of Shannon entropy. As all $p_{ji}^A$ and $p_{0i}^B$ can be regarded
as distinct, the equality in Eq. (\ref{concavity}) holds when there exists only one pure
component, i.e., $p_k=1$ for a certain $k$ and $p_i=0$ for $i\ne k$. Similarly,
$H(A|B)\ge H^{pcs}(A|B)$. Hence, for all separable, or classically correlated, states
$\rho$, $\Xi(\rho) \le \Xi^{pcs}(\rho)$ and thus $\Xi(\rho) \le \Xi(\ket{00})$.
\hfill{$\Box$}

\begin{figure}
\begin{center}
\begin{picture}(80,60)
  \put(5,2){\includegraphics[scale=0.3]{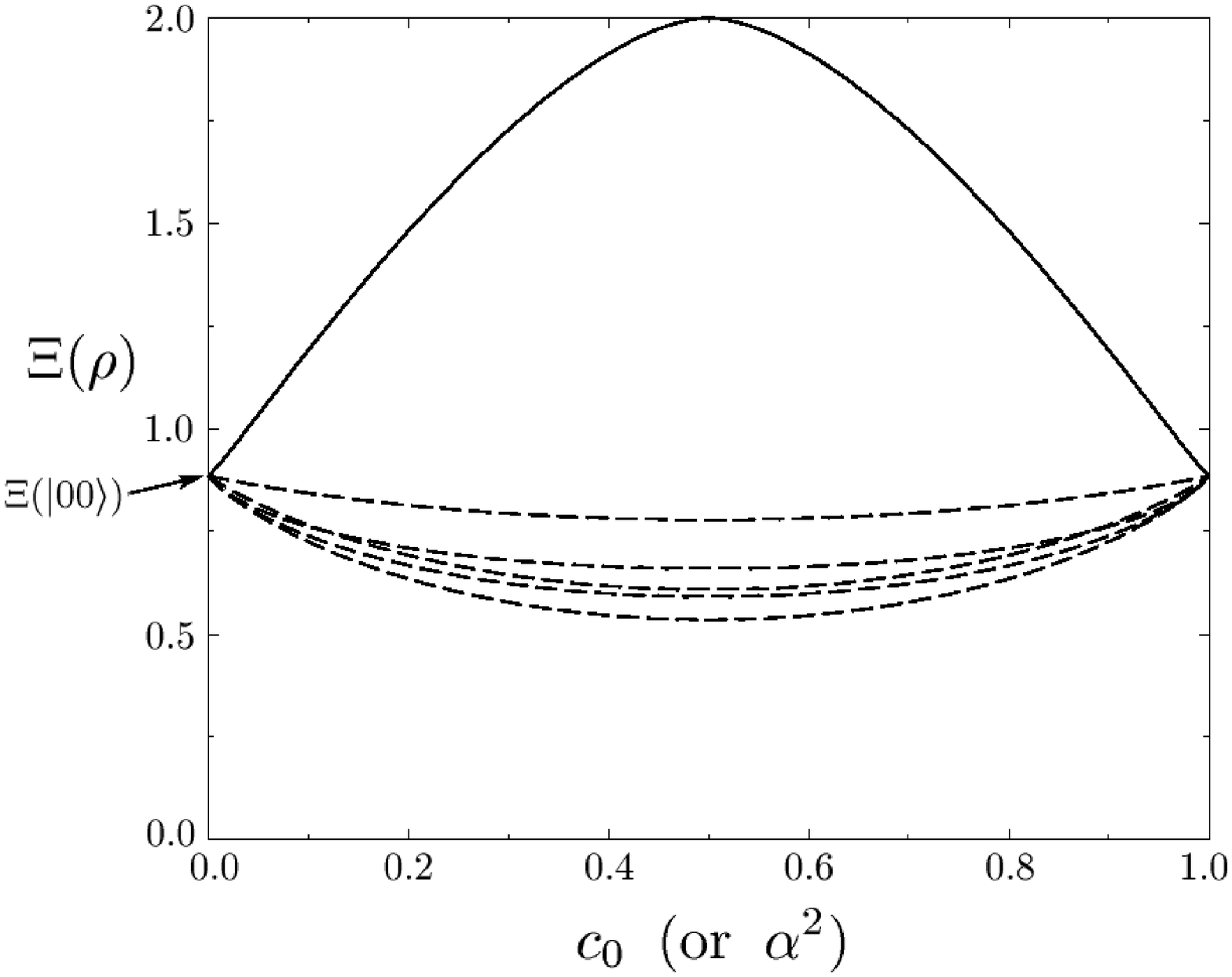}}
\end{picture}
\caption{Extractable work in the asymptotic limit. The dashed lines show the extractable work
from classical correlated states,
$\sigma_\mathrm{cl.}=c_0\ket{00}\bra{00}+c_1\ket{\varphi\varphi}\bra{\varphi\varphi}$,
where $\varphi\in\{0.2\pi, 0.4\pi, \cdots, \pi\}$. The solid line is that from entangled
pure states, $\alpha\ket{00}+\beta\ket{11}$. The horizontal axis represents two
parameters, namely, $c_0$ for classical correlation and $\alpha^2$ for entanglement.}
 \label{ineq_plot}
 \end{center}
\end{figure}

The dashed lines in Fig. \ref{ineq_plot} show numerically plotted $\Xi(\sigma_\mc{cl.})$ as
a function of $c_0$, where
$\sigma_\mc{cl.}=c_0\ket{00}\bra{00}+c_1\ket{\varphi\varphi}\bra{\varphi\varphi}$ is a
classically correlated state with respect to two vectors, $\ket{0}$ and
$\ket{\varphi}:=\cos(\varphi/2)\ket{0}+\sin(\varphi/2)\ket{1}$ There are five dashed
lines, each of which corresponds to $\sigma_\mc{cl.}$ with one $\varphi$ out of the set
$\{0.2\pi, 0.4\pi, \cdots, \pi\}$. The near-horizontal one corresponds to
$\varphi=0.2\pi$, which gives a state ``close'' to $\ket{00}$. We can see that none of
$\Xi(\sigma_\mc{cl.})$ exceeds $\Xi(\ket{00})$, whose value is 0.8854 bits, as the above
proposition claims. Also, above the threshold $\Xi(\ket{00})$, the extractable work from
a (pure) entangled state is a monotonic function of the amount of entanglement
\cite{vedral97}, taking its maximum when maximally entangled.

As the RHS of Eq. (\ref{theineq}) represents the maximum amount of work obtainable from
classically correlated states with our protocol, any excess of extractable work should be
the manifestation of entanglement. Thus, violating the inequality (\ref{theineq}), demons
can exploit the extra work from entanglement, which is unavailable from classically
correlated states.

Even if we choose a part of the great circle as the integral path of Eq.
(\ref{var_basis}) without closing it, the inequality corresponding to Eq. (\ref{theineq})
should be violated for states that are entangled strongly enough. This is because for a
strongly entangled state $\ket{\phi}$, $\xi_{\ket{\phi}}(A_\theta,B_\theta)$ is close to
2 for all $\theta$, while $\xi_{\ket{00}}(A_\theta,B_\theta)$ is always less than 2
unless $\theta$ indicates $\ket{0}$. This also means that the violation criterion, i.e.,
the RHS of Eq. (\ref{theineq}), depends on the range of the path in such a case.

We can also perform the integral in Eq. (\ref{var_basis}) over the whole Bloch sphere,
instead of the great circle, to have another separability criterion.
Then, Eq. (\ref{theineq}) becomes
\begin{equation}\label{bloch}
 \Xi_{BS}(\rho):=\frac{1}{4\pi}\int_{BS}\xi_\rho(A,B)d\Omega\le \Xi_{BS}(\ket{00}),
\end{equation}
where $BS$ stands for the Bloch sphere and we obtained $\Xi_{BS}(\ket{00})=0.5573$
numerically. The proposition also holds for $\Xi_{BS}(\rho)$ as there is no need to
change the proof except that $\Xi_{BS}^{pcs}(\rho)$ is always equal to
$\Xi_{BS}(\ket{00})$ in this case. Let us now compute the value of $\Xi_{BS}(\rho_W)$,
where $\rho_W=p\ket{\Psi^-}\bra{\Psi^-}+(1-p)/4\cdot I$ is the Werner state
\cite{werner89}, to see the extent to which the inequality can be satisfied when we vary
$p$. It has been known that Bell-CHSH inequalities \cite{chsh} are violated for
$p>1/\sqrt{2}=0.7071$, while $\rho_W$ is inseparable iff $p>1/3$. A bit of algebraic
calculations lead to $\Xi_{BS}(\rho_W)=(1-p)\log_2(1-p)+(1+p)\log_2(1+p)$ and this is
greater than $\Xi_{BS}(\ket{00})$ when $p>0.6006$. Therefore, the inequality
(\ref{bloch}) is significantly stronger than Bell-CHSH inequalities.

Since we have obtained the inequalities (\ref{theineq}) and (\ref{bloch}) without
discussing the non-locality of quantum mechanics, they are different from Bell's
inequalities, but similar to them in the sense that they discriminate non-classical
correlations from classical ones. With reference to Bell's inequalities, the form of
extractable work, Eq. (\ref{xidef}), reminds us of the ``information theoretic Bell's
inequalities'' derived by Schumacher by defining the \textit{information distance} using
conditional entropies \cite{schumacher91}.

Schumacher obtained his inequalities from the triangle inequality for a metric. However,
it is possible to derive the same inequalities directly from our definition of the
extractable work, Eq. (\ref{xidef}). The result becomes
\begin{equation}\label{1stineq}
\xi(A,B)+\xi(B,C) \le 2+\xi(A,C),
\end{equation}
with equality when $H(A)=0$ or $H(C)=0$. This is a direct consequence of Eq.
(\ref{xidef}) and the strong subadditivity of the Shannon entropy. From Eq.
(\ref{1stineq}), we can obtain ``chained" inequalities
\begin{eqnarray}\label{chainedineq}
\xi(A_1,B_1)+\xi(B_1,A_2)+\cdots+\xi(A_n,B_n) \nonumber \\
\le 2(2n-2)+\xi(A_1,B_n).
\end{eqnarray}
Note that the left-hand side of Eq. (\ref{chainedineq}) becomes
the same as that of Eq. (\ref{theineq}) if we take bases $A_1,
\cdots, B_n$ as those in Fig. \ref{arrows}, divide the sum by
$2n-1$, and let $n$ tend to infinity. However, the same procedure
on the RHS gives $2$, thus Eq. (\ref{chainedineq}) becomes a
trivial inequality $\Xi(\rho)\le 2$. Hence, our inequality
(\ref{theineq}) is essentially independent of Bell-Schumacher
inequalities and gives a stronger bound on $\Xi(\rho)$ in the
continuous limit.

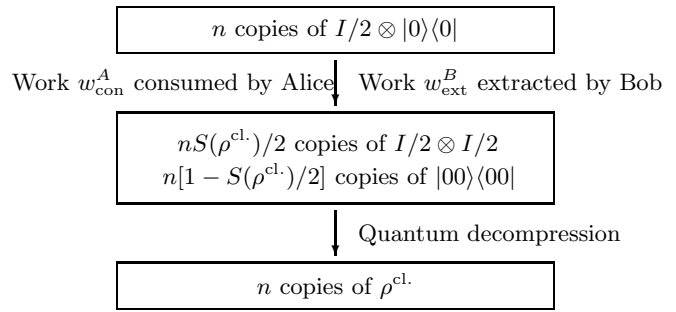
\begin{figure}
\begin{picture}(80,44)
\put(11,35){\framebox(58,6){$n$ copies of $I/2\otimes\ket{0}\bra{0}$}}
\put(40,34){\vector(0,-1){6}} \put(11,15){\framebox(58,12)}
\put(19,22){$nS(\rho^\mc{cl.})/2$ copies of $I/2\otimes I/2$}
\put(17,17.5){$n[1-S(\rho^\mc{cl.})/2]$ copies of $\ket{00}\bra{00}$}
\put(40,14){\vector(0,-1){6}} \put(40,14){\line(0,-1){6}} \put(11,1){\framebox(58,6){$n$
copies of $\rho^\mc{cl.}$}} \put(-3,30){Work $w_\mc{con}^A$ consumed by Alice}
\put(43,30){Work $w_\mc{ext}^B$ extracted by Bob} \put(43,10){Quantum decompression}
\end{picture}
\caption{Restoration of the initial state $\rho^\mc{cl.}$. It is simply a scheme to
prepare $nS(\rho^\mc{cl.})$ of $I/2$ from $n$ copies of the state $I/2\otimes
\ket{0}\bra{0}$. The work, $w_\mc{con}^A$ and $w_\mc{ext}^B$, are explained in the main
text.}\label{decomp}
\end{figure}

\section{Analysis of demons' attempt}
Let us analyze how demons' attempt to violate the Second Law of thermodynamics will end
in failure. In order to discuss the Second Law, which states ``there does not exist any
cycle of a heat engine that converts heat to work without leaving any change in its
environment,'' we need to restore the initial state after extracting work and check the
balance of work. The impossibility of breaking the Second Law by the protocol described
above follows from the fact that the net \textit{work investment} is always nonnegative,
regardless of the direction of measurement/work extraction, thus, it is necessarily
nonnegative after averaging over the great circle or the Bloch sphere.

Suppose that Bob performs measurement on his part and Alice
extracts work on her side. The state after work extraction is
$\sigma=I/2\otimes\ket{0}\bra{0}$. If Bob's outcome was
1, he can flip it with $\sigma_x=\left( \begin{array}{cc} 0 & 1 \\
1 & 0 \end{array} \right)$ without energy consumption. Generically, unitary operations
require no energy consumption as entropy stays constant with them. Let us consider first
the case in which there was only classical correlation initially, as
$\rho^\mc{cl.}=\sum_i p_i \rho_i^A\otimes \rho_i^B$. The simplest method to restore $n$
copies of this initial state from $\sigma$ is to make use of quantum data
compression/decompression \cite{schumacher95}. To do this, they need $nS(\rho^\mc{cl.})$
copies of $I/2$ and $n[1-S(\rho^\mc{cl.})]$ copies of the standard pure state, $\ket{0}$,
where $S(\rho)=-\mc{Tr}\rho\log_2\rho$ is the von Neumann entropy. Let them have the same
number of copies of $I/2$ (See Fig. \ref{decomp}). Alice compresses
$n[1-S(\rho^\mc{cl.})/2]$ copies of $I/2$ on her side isothermally to $\ket{0}$,
consuming $w_\mc{con}^A=1-S(\rho^\mc{cl.})/2$ bits of work per qubit. On the other hand,
Bob acquires $w_\mc{ext}^B=S(\rho^\mc{cl.})/2$ bits of work per qubit by transforming
$nS(\rho^\mc{cl.})/2$ copies of $\ket{0}$ to $I/2$. They can restore $n$ copies of
$\rho^\mc{cl.}$ by decompressing the mixture of $I/2\otimes I/2$ and $\ket{00}$ globally
without any work consumption.

When the initial state is entangled as $\ket{\psi^{AB}}=\alpha\ket{00}+\beta\ket{11}$,
the restoration process becomes simpler. Alice transforms her state $I/2$ into $\ket{0}$
by consuming one bit of energy to make $\ket{00}$. Then rotating Alice's state to
$(\alpha\ket{0}+\beta\ket{1})\ket{0}$ unitarily and applying a global CNOT restore the
initial entanglement. In either case, we also need to erase the information of Bob's
measurement outcome and this requires $H(B(\theta))$ bits of work \cite{demon,landauer}.

Combining the work extracted before, we can now calculate the work investment
$W_\mc{inv}$ to close the cycle:
$W^{\mc{cl.}}_{\mc{inv}}=w_\mc{con}^A-w_\mc{ext}^B+H(B(\theta))
-[1-H(A(\theta)|B(\theta))]=H(A(\theta),B(\theta))-S(\rho^\mc{cl.})$ bits of work for an
initial state of $\rho^\mc{cl.}$ and, similarly,
$W^\mc{ent.}_\mc{inv}=H(A(\theta),B(\theta))$ bits for an entangled state,
$\ket{\psi^{AB}}$. These $W_{\mc{inv}}$ must be nonnegative in order for the Second Law
not to be violated. It turns out that both $W^\mc{cl.}_\mc{inv}$ and
$W^\mc{ent.}_\mc{inv}$ are indeed nonnegative due to the properties of the Shannon and
von Neumann entropies and the effect of projective measurements.

\section{Summary}
We have devised a scenario in which two demons, Alice and Bob, can distill more work from
entanglement than from classical correlation. We have cast this discrimination of
correlations in the form of the thermodynamical separability criteria. Although we can
re-derive Schumacher's inequalities by considering a set of discrete basis of
measurement, our inequalities are essentially different. Interestingly, our inequality
(\ref{bloch}) is more effective than Bell-CHSH inequalities, as well as Schumacher's, in
detecting inseparability of the Werner state. Lastly, our analysis of the energy balance
after closing the thermodynamical cycle illustrates, quite expectedly, that the demons
cannot violate the Second Law even with the excess work extracted from entangled states.

One important step we should make next is to devise a method to test the inequality
experimentally, for example, by using NMR as proposed in Ref. \cite{lloyd97}. Also,
it would be very interesting if we can find a connection between
our inequalities and the non-locality of quantum mechanics.

\section*{Acknowledgments}
KM acknowledges financial support by Fuji Xerox. FM appreciates the warmth and
hospitality of the quantum information group at Imperial College. VV is supported by
the European Community, the Engineering and Physical Sciences Research Council, Hewlett-Packard
and Elsag Spa.


\begin{thebibliography}{99}
\bibitem{bell64}
J. S. Bell, Physics (Long Island City, N.Y.) \textbf{1}, 195 (1964).
%
\bibitem{oppenheim02}
J. Oppenheim, M. Horodecki, P. Horodecki, and R. Horodecki, Phys.
Rev. Lett. \textbf{89}, 180402 (2002).
%
\bibitem{zurek03}
W. H. Zurek, Phys. Rev. A \textbf{67}, 012320 (2003); W. H. Zurek, Rev. Mod. Phys.
\textbf{75}, 715 (2003).
%
\bibitem{terhal00}
B. M. Terhal, Phys. Lett. A \textbf{271}, 319 (2000).
%
\bibitem{werner89}
R. F. Werner, Phys. Rev. A. \textbf{40}, 4277 (1989).
%
\bibitem{szilard29}
L. Szilard, Z. Phys. \textbf{53}, 840--856 (1929)
%
\bibitem{bennett82}
C. H. Bennett, Int. J. Theor. Phys. \textbf{21}, 905 (1982)
%
\bibitem{vonneumann}
J. von Neumann, \textit{Mathematical Foundations of Quantum Mechanics}, (English
translation by R. T. Beyer, Princeton University Press, Princeton, 1955).
%
\bibitem{demon}
H. S. Leff and A. F. Rex (eds.), \textit{Maxwell's Demon 2},
(Institute of Physics Publishing, Bristol and Philadelphia, 2003).
%
\bibitem{braunstein90}
S. L. Braunstein and C. M. Caves, Ann. Phys. \textbf{202}, 22 (1990).
%
\bibitem{note_ent}
Since we are focusing on qubits, the entropy $H(B|A)$ can be determined with a probability
for only one outcome on Bob's side, say 0, for instance. This is the same as taking
$p=\mc{Tr}(P_0\rho)$ for the (binary) Shannon entropy $H(p)$ in Section \ref{sec_wes}.
%
\bibitem{vedral97}
V. Vedral, M. B. Plenio, M. A. Rippin, and P. L. Knight, Phys. Rev. Lett. \textbf{78},
2275 (1997).
%
\bibitem{chsh}
J. F. Clauser, M. A. Horne, A. Shimony, and R. A. Holt, Phys. Rev.
Lett. \textbf{23}, 880 (1969).
%
\bibitem{schumacher91}
B. W. Schumacher, Phys. Rev. A. \textbf{44}, 7047 (1991).
%
\bibitem{schumacher95}
B. Schumacher, Phys. Rev. A. \textbf{51}, 2738 (1995).
%
\bibitem{landauer}
R. Landauer, IBM J. Res. Dev. \textbf{5}, 183 (1961).
%
\bibitem{lloyd97}
S. Lloyd, Phys. Rev. A. \textbf{56}, 3374 (1997).
%
\end{thebibliography}
\end{document}